\documentclass[a4paper,12pt]{article}
\usepackage{amssymb,amsfonts,amsmath,mathtext,cite,enumerate,float}
\usepackage{amssymb,amsthm,latexsym,euscript,amscd, authblk}

\usepackage[russian,english]{babel}
\usepackage[cp1251]{inputenc}

\title{On construction of anticliques for non-commutative operator graphs}

\author[1]{G.G.Amosov\thanks {gramos@mi.ras.ru}}

\author[2]{A.S. Mokeev\thanks {alexandrmokeev@yandex.ru}}

\affil[1]{Steklov Mathematical Institute of Russian Academy of Sciences}
\affil[2]{Saint-Petersburg State University}

\begin{document}

\maketitle

\begin{abstract}
In this paper, we construct anticliques for non-commutative operator graphs generated by generalized Pauli matrices.
It is shown that application of entangled states for the construction of the code space $K$ allows one to substantially increase the dimension of a non-commutative operator graph for which the projection on $K$ is an anticlique.  
\end{abstract}


\section{Introduction}

In \cite{ChoiEffros} the authors introduced operator systems which are subspaces  $\mathcal V$ of the algebra of all bounded linear operators in a Hilbert space $H$ containing the identity operator $I\in \mathcal V$ and such that the following implication holds:
$$
 A\in {\mathcal V}\ \Rightarrow A^*\in {\mathcal V}. 
$$
In the context of quantum information theory, operator systems were called {\it non-commutative operator graphs} \cite{Duan, Weaver}.  We treat operator systems in the spirit of the theory of quantum error-correcting codes  \cite{Knill, Knill2}. 
Graphs consisting of commuting operators are a particular case in this theory.
We assume that $\mathcal V$ is
generated by a set of unitary operators $U_j,\ 1\le j\le k$ called errors, so that 
$$
{\mathcal V}= Lin \{I,U_1,U_1^*,
\dots ,U_k,U_k^*\}.
$$
 {\it A quantum code} is a subspace $K\subset H$. It is assumed that some information is coded in unit vectors $\psi \in H.$  Since a minimal volume of information is one qubit we conclude that $\dim K\ge 2$. 
The dimension of $K$ is said to be {\it a code  lenght}.
Any choice of an orthonormal basis $(f_j)$ in $K$ fixes the set of words $(f_j)$ which can be used to code quantum or classical information. The code $K$ is said to be {\it a quantum error-correcting code} for $\mathcal V$ if 
\begin{equation}\label{anticlique}
\dim P_{K}{\mathcal V}P_K=1,
\end{equation}
where $P_K$ is an orthogonal projection onto the subspace $ K $. 
By definition, $I\in {\mathcal V}$, therefore it follows from (\ref {anticlique}) that
$$
P_KVP_K=c_{V}P_K,\ c_V\in {\mathbb C},
$$
for each $V\in {\mathcal V}$. Hence, for the orthonormal basis $(f_j)$ in $K$ we have the relations
$$
(f_j,Vf_k)=\delta _{jk}c_V.
$$
By this way, vectors $Vf_j$ and $f_k$ are orthogonal, if $j\neq k$  and  $V\in {\mathcal V}$.
Thus, after the implementation of errors from $ \mathcal V $ on the code $K$, we get new words which could be perfectly distinguished.  This motivates one to call codes with this properties {\it quantum error-correcting codes}. In classical graph theory the set of $ k $ vertices of the graph with $ n $ vertices is said to be an ancticlique if vertices from this set are not pairwise connected by edges.
Similarly, the projections $P_K$ satisfying (\ref{anticlique})
were called in \cite{Weaver2} {\it quantum anticliques}.
If a noncommutative operator graph $\mathcal V$ is the maximal commutative algebra on a Hilbert space $H$, then there are no quantum anticliques for $\mathcal V$ \cite{Weaver2}.

Our goal is to construct non-commutative operator graphs $\mathcal V$ for which there exist anticliques. 
Previously, such problem was solved for abstract non-commutative operator graphs. 
In \cite{Knill2} it was proved that in a space $H,\ \dim H=n,$ there is an anticlique $P_K,
\dim K=k$, 
if the following condition holds
\begin{equation}\label{est+}
\dim {\mathcal V}( \dim {\mathcal V}+1) \le \frac {n}{k}.
\end{equation}
The proof of this fact is not constructive and based upon deep combinatorial results  \cite {Tverberg1, Tverberg2}. In a particular case where the graph is generated by a set of commuting operators, bound (\ref {est+}) can be replaced by 
\begin{equation}\label{est}
\dim {\mathcal V}\le \frac {n-k}{k-1}
\end{equation}
(\cite {Weaver}).
In \cite {Knill, Knill2} and other papers of this school the problem of constructing anticliques was solved for the graphs generated by Pauli matrices. In contrast to our paper, codes for a tensor product of $ N $ copies of the two-dimensional space were considered.

{\it A quantum state} is a positive operator with
unit trace on a Hilbert space $H$ corresponding to quantum system. The set of all states is denoted by $\mathfrak {S}(H)$. An important role is played by pure states, i.e. orthogonal projections too one-dimensional spaces.
Composite quantum systems are described by the states from $\mathfrak {S}(H\otimes K)$ on the tensor product of the Hilbert spaces $H$ and $K$. In this case, we must distinguish {\it separable states} and  {\it entangled states.}  Separable states can be represented as $\rho =\sum \limits _j\pi _j\rho _j\otimes \tilde \rho _j,$ where $\rho _j\in \mathfrak {S}(H)$, $\tilde \rho _j\in \mathfrak {S}(K)$, $\pi_j \ge 0$ and $\sum \limits _j\pi _j=1.$  Entangled states do not have this property. The presence of entangled states is a purely quantum effect and their use for
coding can give an improvement in some important cases \cite{Shor}. The main purpose of the proposed work is to estimate the benefits given by application of entangled states to build quantum anticliques. Since the notion of entanglement is defined in the spaces $H$ which are tensor products of two spaces the problem makes a sense only for dimensions $ \dim H \ge 4 $ (the minimal dimension corresponds to a tensor product of two two-dimensional spaces).

In this paper we construct non-commutative operator graphs $\mathcal V$ in a Hilbert space $H$ that have anticliques $P_K$
onto subspaces $K\subset H$. In the second section, $\dim H=4,\ \dim {\mathcal V}=4$ and $\dim K=2$. In the third one, $\dim H=n^2,\ \dim {\mathcal V}=n(n-1)+1$ and
$\dim K=n$. In these sections the subspace $K$ is generated by separable states. In the fourth section, we show that application of entangled states in the construction of anticliques allows us to increase the dimension of the graph $\mathcal V$ consisting of correctable errors.

We are interested in the case where the number of correctable errors and the quantity of information protected by coding increase together with the dimension of the space $H$. The large dimension of errors gives one a chance to work on a low level of protection against decoherence. This problem is the main difficulty in the design of quantum information systems \cite{Shor}. Let us consider $\dim H=n^2$ (a tensor product of two $n$-dimensional spaces). If, for example, we want the dimension of anticlique to be $n$, then Estimate $(\ref{est+})$ leads to the upper estimate $O(\sqrt n)$ for the dimension of the operator graph, and Estimate (\ref {est}) leads to the estimate $O(n)$ for the commutative case.
 In our examples, the dimension of graphs has order $O(n^4)$ under the same conditions on the dimension of anticliques.

\section{Error-correcting codes in a four-dimensional space.}

It follows from (\ref {est}) that an arbitrary non-commutative
operator graph ${\mathcal V}$, $dim {\mathcal V}=2$ in a Hilbert space ${\mathcal H}$ with $\dim {\mathcal H}=4$  has an anticlique
$P_K$ with dimension $\dim K=2$. The construction of such system is really simple. Let ${\mathcal H}={\mathbb C}^2\otimes
{\mathbb C}^2$, $\sigma _x=\begin{pmatrix}0&1\\ 1&0 \end{pmatrix},\sigma _y=\begin{pmatrix}0&i\\-i&0 \end{pmatrix}$, $\sigma_z=\begin{pmatrix}1&0\\0&-1 \end{pmatrix}$ be
the standard Pauli matrices. Assume that $\mathcal V$ is generated by $I\otimes \sigma _x$. Then the projection $P_K$ onto the subspace
$K=\left \{{\mathbb C}^2\otimes \begin{pmatrix}\lambda\\0\end{pmatrix},\ \lambda \in {\mathbb C}\right \}$, $\dim K=2$, is an anticlique for ${\mathcal V}=Lin\{I\otimes I,\  
I\otimes \sigma _x\}$

Later we construct a graph ${\mathcal V},\ \dim {\mathcal V}=3,$ which have an anticlique  $P_K,\ \dim K=2$. In addition,
we show that we can add two additional elements to $\mathcal V$, so that we get $\dim {\mathcal V} = 5 $ and $P_K$ remains an anticlique for the extended non-commutative graph as well.

Consider the matrices $T=\sigma _x\otimes I,\ U=\sigma _y\otimes I$, $V=I\otimes \sigma _y$ and $W=I\otimes \sigma _z$ in ${\mathcal H}={\mathbb C}^2\otimes {\mathbb C}^2$. 
Let us define vectors $f_{\pm}\in {\mathcal H}$ by the formulae
$$
f_+=\begin{pmatrix}1\\0\end{pmatrix}\otimes \begin{pmatrix}1\\1\end{pmatrix},\ f_-=\begin{pmatrix}0\\1\end{pmatrix}\otimes \begin{pmatrix}1\\-1\end{pmatrix}.
$$

{\bf Proposition 1.} {\it We have the equalities
$$
(f_{\pm },xf_{\pm})=0,\ (f_{\pm},xf_{\mp})=0,
$$
$x\in \{T,U,V,W\}$.
}

Proof.

We prove this proposition for $x=U$. The remaining cases can be treated analogously.
Note that
$$
(f_{+},Uf_{+})=\left (\begin{pmatrix}1\\0\end{pmatrix}\otimes \begin{pmatrix}1\\1\end{pmatrix},\begin{pmatrix}0\\i\end{pmatrix}\otimes \begin{pmatrix}1\\1\end{pmatrix}\right )=0,
$$
$$
(f_{-},Uf_{-})=\left (\begin{pmatrix}0\\1\end{pmatrix}\otimes \begin{pmatrix}1\\-1\end{pmatrix},\begin{pmatrix} i \\ 0\end{pmatrix}\otimes \begin{pmatrix}1\\-1\end{pmatrix}\right )=0
$$
and
$$
(f_{+},Uf_{-})=\left (\begin{pmatrix}1\\0\end{pmatrix}\otimes \begin{pmatrix}1\\1\end{pmatrix},\begin{pmatrix}i\\0\end{pmatrix}\otimes \begin{pmatrix}1\\-1\end{pmatrix}\right )=0,
$$

$\Box$

Consider the subspace $K,\ \dim K=2$, generated by $f_+$ and $f_-$.

{\bf Corollary.} {\it We have the equality
$$
\dim P_K{\mathcal V}P_K=1.
$$
}

\pagebreak

Proof.

It follows from Proposition 1 that $P_KxP_K=0$ if $x\in \{U,V,W\}$. By the definition of a non-commutative operator graph, in addition to $U,V,W$  the identity operator $I$ belongs to $\mathcal V$ as well.
It remains to note that $P_KIP_K=P_K$. $\Box $

{\bf Remark 1.} {\it The constructed non-commutative operator graph and the quantum anticlique correspond to the Shor's scheme for correcting errors \cite{Shor}.
But in the Shor scheme a correction is possible with asymptotically zero error. In our schem we have zero-error correction. Nevertheless we suppose that only two error (for example $\sigma _x$ and $\sigma _y$) from three possible errors $\sigma_x,\sigma_y,\sigma_z$ can occur on a certain qubit .}

\section{Spaces of dimension $n^2$}

In this section we generalize the construction to the spaces ${\mathcal H}={\mathbb C}^{n}\otimes {\mathbb C}^n,\ n>2$, such that $\dim {\mathcal H}=n^2$. Let $(e_j)_{j=1}^{n}$ be the standard basis in
${\mathbb C}^n$. Consider a complex Hadamard matrix $H=(a_{jk})_{j,k=1}^n$ with the coefficients
$a_{jk}=e^{\frac {2\pi i}{n}(j-1)(k-1)},\ 1\le j,k\le n$. It is known that \cite {Haagerup}, the columns of this matrix
$$
f_j=\frac {1}{\sqrt n}\sum \limits _{k=1}^na_{jk}e_k,\ 1\le j\le n,
$$
form an orthonormal basis of the space ${\mathbb C}^n$.
Consider unitary operators $X$ and $Z$ that act on ${\mathbb C}^n$ by the formulae
$$
Xe_j=e^{\frac {2\pi i}{n}(j-1)}e_j,\ Zf_j=e^{\frac {2\pi i}{n}(j-1)}f_j,
$$
$1\le j\le n$.
Note that
\begin{equation}\label{shift}
Xf_j=f_{j+1\ mod\ n},\ 1\le j\le n.
\end{equation}
The operators $Y_{km}=X^kZ^m,\ 0\le k,m\le n-1$, which are called generalized Pauli matrices, satisfy the Heisenberg-Weyl relations
\begin{equation}\label{HW}
Y_{km}Y_{k'm'}=e^{\frac {2\pi i}{n}(mk'-km')}Y_{k'm'}Y_{km},\ 0\le ,m,k,m',k'\le n-1.
\end{equation}

Let us introduce vectors $h_j\in {\mathcal H}$
by the formulae
$$
h_j=f_j\otimes f_j,
$$
$1\le j\le n$.

Denote $U_k=XZ^k\otimes I, V_k=I\otimes XZ^k$,
$0\le k\le n-1$

{\bf Proposition 2.} {\it The equalities
$$
(h_j,U_k^sh_m)=(h_j,V_k^sh_m)=0
$$
are valid for $1\le j,m\le n,\ 1\le s\le n-1 ,\ 0\le k\le n-1$.}

Proof.

Let us prove Proposition for operators $U_k$. The case of $V_k$ is analogous.
We note that 
$$ 
(h_j,U_k^sh_k)= (f_j\otimes f_j,((XZ^k)^s\otimes I)(f_k\otimes f_k))=0
$$
if $j\neq k$ and
$$
(h_j,U_k^sh_j)=(f_j,(XZ^k)^sf_j).
$$
From (\ref {HW}) we get $(XZ^k)^s=c_{ks}X^sZ^k,$
where $|c_{ks}|=1$. By this way,
$$
(f_j,(XZ^k)^sf_j)=\overline {c_{ks}}e^{-\frac {2 \pi i}{n}ksj}(f_j.f_{j+s\ mod\ n})=0,
$$
because $s\neq 0\ mod\ n. $

$\Box $

Let a graph $\mathcal V$ be generated by the operators
$U_k^s,V_k^s,\ 0\le k\le n-1,\ 1\le s\le n-1$. Denote $P_K$ 
a projection onto the $k$-dimensional subspace $K$ generated by the vectors $h_j,\ 1\le j\le n$.

{\bf Theorem 1.} {\it $P_K$ is a quantum anticlique for  ${\mathcal V}$.}

Proof.

It follows from Proposition 2 that $P_KU_k^sP_K=P_KV_k^sP_K=0,\ 0\le k\le n-1,\ 
1\le s\le n-1$. For a fixed $k$ the operators $(U_k^s)_{s=0}^{n-1}$ , $(V_k^s)_{s=0}^{n-1}$
form the cyclic groups. Thus, 
\begin{equation}\label{oper}
{\mathcal V}=Lin(U_k^s,V_k^s,(U_k^s)^{*},(V_k^s)^{*}, 0\le k,s\le n-1).
\end{equation}
and $P_K$ is an anticlique.

$\Box $

{\bf Theorem 2.} {\it $\dim {\mathcal V}=2n(n-1)+1$.}

Proof.

Note that $Z$ and $X$ are unitary operators. Moreover, $Z$ and $X$ generate cyclic groups of order $n$. Using the Heisenberg-Weyl relations 
(\ref{HW}) we get
$$
((XZ^k)^s)^*= cX^{-s}Z^{-ks}=cX^{n-s}Z^{-ks}=cX^{j}Z^{m}, \  1\le j \le n-1, \  0 \le m \le n-1, \ 
$$
$c \in \mathbb{C}.$  
It implies that
$$
Lin((U_k^s)^{*},(V_k^s)^{*}, 0\le k,s\le n-1) \subset Lin(U_k^s,V_k^s 0\le k,s\le n-1).
$$
Hence, we can restrict the generating set of graph ${\mathcal V}=Lin(U_k^s,V_k^s 0\le k,s\le n-1)$.
Now the result follows from a linear independence of the generalized Pauli matrices.

$\Box $

\section{Application of entangled states for a construction of anticliques}

In this section we show that the non-commutative operator graph considered in the previous section can be expanded to a new graph that has also an anticlique. This is achieved by application of entangled states.

Let us define a map $\tau:\mathbb{Z}_n \rightarrow \mathbb{N}$ by the formula $\tau(j)=j+1$. This map helps us to parametrize the vectors $f_{j}$ 
by elements of the cyclic group $\mathbb{Z}_n$. Let the dimension $n=p \cdot y$, where $p,y \in \mathbb{N}$. Then $\mathbb{Z}_p$ is a cyclic subgroup in $\mathbb{Z}_n$. Let $\chi:\mathbb{Z}_n \rightarrow \{ 0,1 \}$ be the indicator function of $\mathbb{Z}_p$ as a subset of $\mathbb{Z}_n$.
Define a vector by the following formula
\begin{equation}\label{q1}
q_1=\sum_{j=0}^{n-1} \chi(j) f_{\tau(j)} \otimes f_{\tau(j)},
\end{equation}
where the index $j \in  \mathbb{Z}_n$.
Suppose that 
$$(h+1)(d+1) \geq y \geq (h+1)d$$ for two fixed indeces $h,d \in \mathbb{N}$.
Then vector (\ref {q1}) recursively generates the following set of orthogonal vectors $(q_k)_{k=1}^{d}$
$$q_{k+1}=(X^{h+1} \otimes X^{h+1})q_{k},$$
where $1 \leq k < d$. Let $K_{1} = Lin (q_k)_{k=1}^{d} \subset K$ and let $P_{K_{1}}$ be the orthogonal projection onto $K_{1}$.
Here $\left \lfloor \frac{m}{y} \right \rfloor$ denotes the reminder of division by $y$. Let us introduce the set 
$$
A=\{ m \in \mathbb{N}: \left \lfloor \frac{m}{y} \right \rfloor \neq  (d-j)( h+1), \  \left \lfloor \frac{m}{y} \right \rfloor \neq y+(j-d)( h+1) , \ \textit{where}\ 1\le j \le d \}
$$
Consider the following sets of operators
$${\mathcal A}=Lin(X^{m}Z^{k} \otimes X^{j}Z^{s},
(X^{m}Z^{k} \otimes X^{j}Z^{s})^{*}, \ \ m \neq j)$$
$${\mathcal B}=Lin(X^{m}Z^{k} \otimes X^{m}Z^{s}, 
(X^{m}Z^{k} \otimes X^{m}Z^{s})^{*}, \  m \in A)$$
$${\mathcal C}=Lin(X^{m}Z^{k} \otimes X^{m}Z^{s},(X^{m}Z^{k} \otimes X^{m}Z^{s})^{*}, \ k+s \not \equiv 0 \ mod \ p, \ 0\leq m \leq n-1)$$
Let us expand the operator graph $\mathcal V$ (\ref {oper}) by this sets ${\mathcal V}_{1}=Lin({\mathcal A} \cup {\mathcal B} \cup {\mathcal C} \cup {\mathcal V})$.

{\bf Theorem 3.} {\it The projection $P_{K_{1}}$ is a quantum anticlique for  ${\mathcal V}_{1}$.}

Proof.
We have to prove a statement similar to Proposition 2 for
operators from the sets ${\mathcal A},  {\mathcal B}, {\mathcal C}$.

Each operator $V \in {\mathcal A}$ acts from $K$ to $Lin(f_{k} \otimes f_{j}, j \neq k)$, it implies that the required condition is
valid.

Notice, that
$$
(X^{m} \otimes X^{m})q_k=(X^{\left \lfloor \frac{m}{y} \right \rfloor} \otimes X^{\left \lfloor \frac{m}{y} \right \rfloor})q_k
$$
if $ m \in A$. Then it follows from the structure of $A$ we can conclude that vectors $(X^{m} \otimes X^{m})q_k$ and $q_j$ are always distinguished. Thus, these vectors are orthogonal, it results in the required condition for ${\mathcal B}$.

Afterthat,
$$(Z^{k} \otimes Z^{s}q_j,q_l)=0, \textit{where} \ j \neq l.$$
Now it suffices to consider the expression 
$$(Z^{k} \otimes Z^{s}q_j,q_j)=(I \otimes Z^{s+k}q_j,q_j)=$$
$$e^{\frac{2 \pi i}{n}(h+1)(j-1)}e^{-\frac{2 \pi i}{n}(h+1)(j-1)}(I \otimes Z^{s+k}q_1,q_1)=0.$$
The first equality follows from the structural properties of $K$. As $\ k+s \not \equiv 0 \ mod \ p$, then $I \otimes Z^{s+k}q_1$ and $q_1$ are elements of the orthogonal basis in a subspace of dimension $p$  \cite {Haagerup}.
Nothing is essentially changed if we add $X^{m}$ to each component of the tensor product, because the commutative relations 
imply that
$$
X^{m}Z^{k} \otimes X^{m}Z^{s}=c(Z^{k} \otimes Z^{s})(X^{m}\otimes X^{m})
$$
for some $c\in \mathbb {C}$. Thus, for arbitrary $1 \leq l,j \leq d$
$$
((Z^{k} \otimes Z^{s})(X^{m}\otimes X^{m})q_l,q_j)=g(I \otimes Z^{s+k}q_1,q_1)=0,
$$
where $g$ equals zero if $(X^{m}\otimes X^{m})q_l \neq q_j$ or  $g$ is some unimodular factor.
Proof of the orthogonality of these vectors completes the proof of the theorem.

$\Box $

{\bf Remark 2.} {\it It follows form the proof that  $P_K$ is an anticlique for $Lin({\mathcal A} \cup I)$.} 

{\bf Theorem 4.} {\it The dimension of the graph can be found by the formula
$$\dim {\mathcal V}_1=n^{3}(n-1)+(\# A^{`})n^2+(n- \# A^{`})\left ( \frac{y(p-1)(p+2)}{2}+\frac{n (y - 1) }{2}\right )+1,$$ where $\#$ denotes the number of elements of $A^{`}=A \cap \{ j: 1\le j <n\}.$}

Proof.

The same reasoning as in the proof of Theorem 2 shows that
 $$
 (X^{m_1}Z^{k_1} \otimes X^{j_1}Z^{s_1})^{*}=cX^{n-m_1}Z^{k'} \otimes X^{n-j_1}Z^{s'}, c \in \mathbb{C}.
 $$
 Thus, $(X^{m_1}Z^{k_1} \otimes X^{j_1}Z^{s_1})^{*} \in Lin(X^{m}Z^{k} \otimes X^{j}Z^{s},\  m \neq j)$.
It means that $\dim {\mathcal A}=n^{3}(n-1)$ since the number of appiarences of indices $m$ and $j$ equals the number
of permutations of length $2$ in a set of $n$ elements,
which is $n(n-1)$, and the factor $2$ appears from the number of ways to select indices  $k$ and $s$.
 
The generalized Pauli matrices form orthogonal system with respect to to the Hilbert-Schmidt scalar product $(A,B)=tr(AB^{*})$. Thus, to calculate the dimension of $\mathcal{V}_1$ we have to find out whether there exist coinciding generators of $\mathcal{V}_1$ .

For $1 \leq m_1 \leq (n-1),\ m_1 \in A$, let us divide $m_1$ by $y$ with remainder $r$
$$m_1=ty+r,$$
then for $n-m_1$ we have the equality
$$
n-m_1=py-ty-r=(p-t-1)y+(y-r). 
$$
Thus, the system of inequalities

$$
\begin{cases}
 & \left \lfloor \frac{m_1}{y} \right \rfloor \neq (d-j)(h+1),\   \ 0\leq j \leq d \\
 & \left \lfloor \frac{m_1}{y} \right \rfloor \neq  y+(j-d)(h+1), \  \ 0\leq j\leq d
\end{cases}
$$

is equivalent to 
$$
\begin{cases}
& \left \lfloor \frac{n-m_1}{y} \right \rfloor \neq (d-j)(h+1),\   \ 0\leq j \leq d \\
 & \left \lfloor \frac{n-m_1}{y} \right \rfloor \neq  y+(j-d)(h+1), \  \ 0\leq j\leq d.
\end{cases}
$$

It implies that

$$
 Lin((X^{m_1}Z^{k_1} \otimes X^{m_1}Z^{s_1})^{*})=
 $$
 
 $$
 Lin({X^{n-m_1}Z^{k_2} \otimes X^{n-m_1}Z^{s_2}}) \subset Lin(X^{m}Z^{k} \otimes X^{m}Z^{s}, \ m \in A).
 $$

 Hence ${\mathcal B}$ has $(\# A^{`})n^2$ linearly independent generators.
 Analogously,  $ k+s \neq ph_0$ if and only if $k+s \neq p(y-h_0)$, where $1 \le h_0 \le y$.
 Thus, for an arbitrary $ k+s \neq ph$ we have
 $$ Lin((Z^{k} \otimes Z^{s})^{*}) \subset Lin(Z^{k} \otimes Z^{s}, \  k+s \neq ph).$$
 The number of ways to select an operator $Z^{k} \otimes Z^{s}$ so that $k+s=r$, where $0<r<p$, equals $r+1$. It
implies that the number of generators of ${\mathcal C}$ that do not belong to ${\mathcal B}$ can be calculated as
 follows
$$
R_A \sum\limits_{k=0}^{y-1} \sum\limits_{j=1}^{p-1} (j+1+kp)=R_A\left ( \frac{y(p-1)(p+2)}{2}+\frac{p (y - 1) y}{2}\right ), 
$$ 
where $R_A=(n- \# A^{`})$.
From this we obtain that 
$$\dim {\mathcal V}_1=n^{3}(n-1)+(\# A^{`})n^2+(n- \# A^{`})\left (\frac{y(p-1)(p+2)}{2}+\frac{n (y - 1) }{2}\right )+1$$
because the identity operator is one more generator.
 
 $\Box $
 
 This work is supported by the Russian Science Foundation under grant No 14-21-00162.

\end{document}